%% file: main.tex
\documentclass[journal]{IEEEtran}

\usepackage[utf8]{inputenc} 
\usepackage[T1]{fontenc}    
\usepackage[hidelinks]{hyperref}       
\usepackage{url}            
\usepackage{booktabs}       
\usepackage{amsfonts}       
\usepackage{nicefrac}       
\usepackage{microtype}      
\usepackage{lipsum}
\usepackage{graphicx}       
\usepackage[numbers, 
sort&compress
]{natbib}

\usepackage{caption}
\usepackage{tikz}
\usepackage{textcomp}
\usepackage{hyperref}
\usepackage{lipsum}

\usepackage{fancyhdr} 
\newcommand\copyrighttext{%
  \footnotesize \textcopyright 2025 IEEE. All rights reserved, including rights for text and data mining and training of artificial intelligence and similar technologies. Personal use is permitted, but republication/redistribution requires IEEE permission. See \href{https://www.ieee.org/publications/rights/index.html} for more information. DOI: \href{https://ieeexplore.ieee.org/document/10903196}{10.1109/JBHI.2025.3543686}}
\newcommand\copyrightnotice{%
\begin{tikzpicture}[remember picture,overlay]
\node[anchor=south,yshift=0pt] at (current page.south) {\fbox{\parbox{\dimexpr\textwidth-\fboxsep-\fboxrule\relax}{\copyrighttext}}};
\end{tikzpicture}%
}

\usepackage{xcolor}
\usepackage{subcaption}
\usepackage{enumitem}
\usepackage{comment}
\usepackage[labelfont=bf]{caption}
\usepackage{multirow}
\usepackage{multicol}
\usepackage{placeins}
\usepackage[separator = {;}]{credits}
\definecolor{another_blue}{rgb}{0.0, 0.47, 0.75}
\definecolor{navyblue}{rgb}{0.0, 0.0, 0.5}
\definecolor{mygreen}{rgb}{0.2,0.7,0.2}
\definecolor{royalblue}{rgb}{0.04,0.33,0.64 }
\usepackage{tabularx}

\hypersetup{
    colorlinks,
    citecolor=black, 
    filecolor=black, 
    linkcolor=black, 
    urlcolor=black, 
}
\usepackage{arydshln}
\graphicspath{{media/}}     

\newcommand{\todo}[1]{\textcolor{red}{TODO: #1}}

\newcommand{\setmincitenames}[1]{%
  \numdef\blx@mincitenames{#1}}
\makeatother

\begin{document}


\title{
COVID-BLUeS - A Prospective Study on the Value of AI in Lung Ultrasound Analysis
}

\author{
Nina Wiedemann, 
Dianne de Korte-de Boer, 
Matthias Richter, 
Sjors van de Weijer, 
Charlotte Buhre, 
Franz A. M. Eggert, 
Sophie Aarnoudse, 
Lotte Grevendonk, 
Steffen Röber, 
Carlijn M.E. Remie, 
Wolfgang Buhre, 
Ronald Henry, 
Jannis Born 



\thanks{
N. Wiedemann is with the Department of Civil, Environmental and Geomatic Engineering, ETH Zurich, Switzerland, e-mail: nwiedemann@ethz.ch
}
\thanks{
D. de Korte-de Boer, C. M.E. Remie, W. Buhre and L. Grevendonk were with the Department of Anesthesiology \& Pain Medicine, Maastricht UMC+, Maastricht, the Netherlands
}
\thanks{
M. Richter and S. Röber were with the Institute of Cognitive Science, University of Osnabrück, Germany
}
\thanks{
S. van de Weijer is with the Center for Acute and Critical Care, Maastricht UMC+, Maastricht, the Netherlands
}
\thanks{
C. Buhre and F.A.M. Eggert are with the Faculty of Health Sciences (FGW), Joint Faculty of the University of Potsdam, the Brandenburg Medical School Theodor Fontane and the Brandenburg Technical University Cottbus-Senftenberg, Germany
}
\thanks{
F.A.M. Eggert was with the Department of Neurosurgery, School for Mental Health and Neuroscience, Maastricht University, the Netherlands
}
\thanks{
S. Aarnoudse and R. Henry are with the Centre for Acute and Critical Care, Emergency Department, Centre for Chronic Diseases, Department of Medicine, Maastricht UMC+, Maastricht, the Netherlands}
\thanks{
W. Buhre is with the Department of Anaesthesiology and Division of Vital Functions, University Medical Center Utrecht, Utrecht, the Netherlands} 
\thanks{
J. Born was with the Department of Biosystems Science and Engineering, ETH Zurich, Switzerland, e-mail: jannis.born@alumni.ethz.ch
}
}

\maketitle

\thispagestyle{fancy}
\copyrightnotice

\vspace{-1em}

\begin{abstract}
As a lightweight and non-invasive imaging technique, lung ultrasound (LUS) has gained importance for assessing lung pathologies. 
The use of Artificial intelligence (AI) in medical decision support systems is promising due to the time- and expertise-intensive interpretation, however, due to the poor quality of existing data used for training AI models, their usability for real-world applications remains unclear.

\textbf{Methods}: 
In a prospective study, we analyze data from 63 COVID-19 suspects (33 positive) collected at Maastricht University Medical Centre.
Ultrasound recordings at six body locations were acquired following the BLUE protocol and manually labeled for severity of lung involvement. 
Anamnesis and complete blood count (CBC) analyses were conducted. 
Several AI models were applied and trained for detection and severity of pulmonary infection.

\textbf{Results:} 
The severity of the lung infection, as assigned by human annotators based on the LUS videos, is not significantly different between COVID-19 positive and negative patients ($p = 0.89$). 
Nevertheless, the predictions of image-based AI models identify a COVID-19 infection with 65\% accuracy when applied zero-shot (i.e., trained on other datasets), and up to 79\% with targeted training, whereas the accuracy based on human annotations is at most 65\%. Multi-modal models combining images and CBC improve significantly over image-only models. 

\textbf{Conclusion}: 
Although our analysis generally supports the value of AI in LUS assessment, the evaluated models fall short of the performance expected from previous work. 
We find this is due to 1) the heterogeneity of LUS datasets, limiting the generalization ability to new data, 2) the frame-based processing of AI models ignoring video-level information, and
3) lack of work on multi-modal models that can extract the most relevant information from video-, image- and variable-based inputs.

To aid future research, we publish the dataset at:~\url{https://github.com/NinaWie/COVID-BLUES}.

\end{abstract}


\section{Introduction}

The COVID-19 pandemic has triggered enormous interest of the machine learning community for medical applications. While PCR tests have quickly become the standard approach to test for a COVID-19 infection, medical imaging has been widely used in clinical practice to assess disease progression, for example, by observing COVID-induced lung damage with Computer Tomography (CT) images. 
An analysis of the publications in 2020 showed that the vast amount of the Machine Learning (ML) literature focuses on X-Ray and CT imaging, whereas 
Lung ultrasound (LUS) has been largely disregarded by the ML community~\citep{born2021role}. 
Indeed, LUS has been recommended for clinical practice since the beginning of the pandemic~\citep{buonsenso2020covid}. In addition to its practical advantages (non-irradiating, hand-held devices), many studies have found a higher sensitivity for LUS compared to chest X-ray \cite{BOURCIER2014115, NAZERIAN2015620, amatya-2018} and several studies comparing LUS to CT scans revealed good agreement and suggested ultrasound as a promising diagnostic tool to identify COVID-19 induced pneumonia~\cite{yang2020lung,wang-2021}. 

One main reason for the slower development of AI methods for ultrasound is the lack of available open-source datasets. 
The largest public dataset was assembled by~\citet{born2021accelerating} and 
includes more than 200 videos 
labelled as one of three classes: viral pneumonia, healthy or bacterial pneumonia. On the other hand, the ICLUS dataset provides US videos with severity score labels~\citep{roy2020deep}. These works have triggered efforts of the machine learning community to improve methods for automated lung ultrasound analysis, targeting  diagnosis~\citep{daveProspectiveRealTimeValidation2023, baumImageQualityAssessment2021, naveenkumarLungUltrasoundCOVID192022, born2021accelerating,robertsUltrasoundDiagnosisCOVID192020,arntfieldDevelopmentConvolutionalNeural2021, macleanCOVIDNetUSTailored2021}, severity assessment~\citep{roy2020deep, lasalviaDeepLearningLung2021}, landmark detection (i.e., detecting pathological patterns in images)~\citep{moshaveghNovelAutomaticDetection2016, brusascoQuantitativeLungUltrasonography2019, roy2020deep,lucassenDeepLearningDetection2023, nicole-2021, wang-2022, mathewsUnsupervisedMultilatentSpace2021, custodeMultiobjectiveAutomaticAnalysis2023} or segmentation~\citep{chengTransferLearningUNet2021,gareDensePixelLabelingReverseTransfer2021,panickerApproachPhysicsInformed2021, roy2020deep}. 

\begin{figure*}[!htb]
    \centering
    \includegraphics[width=1\textwidth]{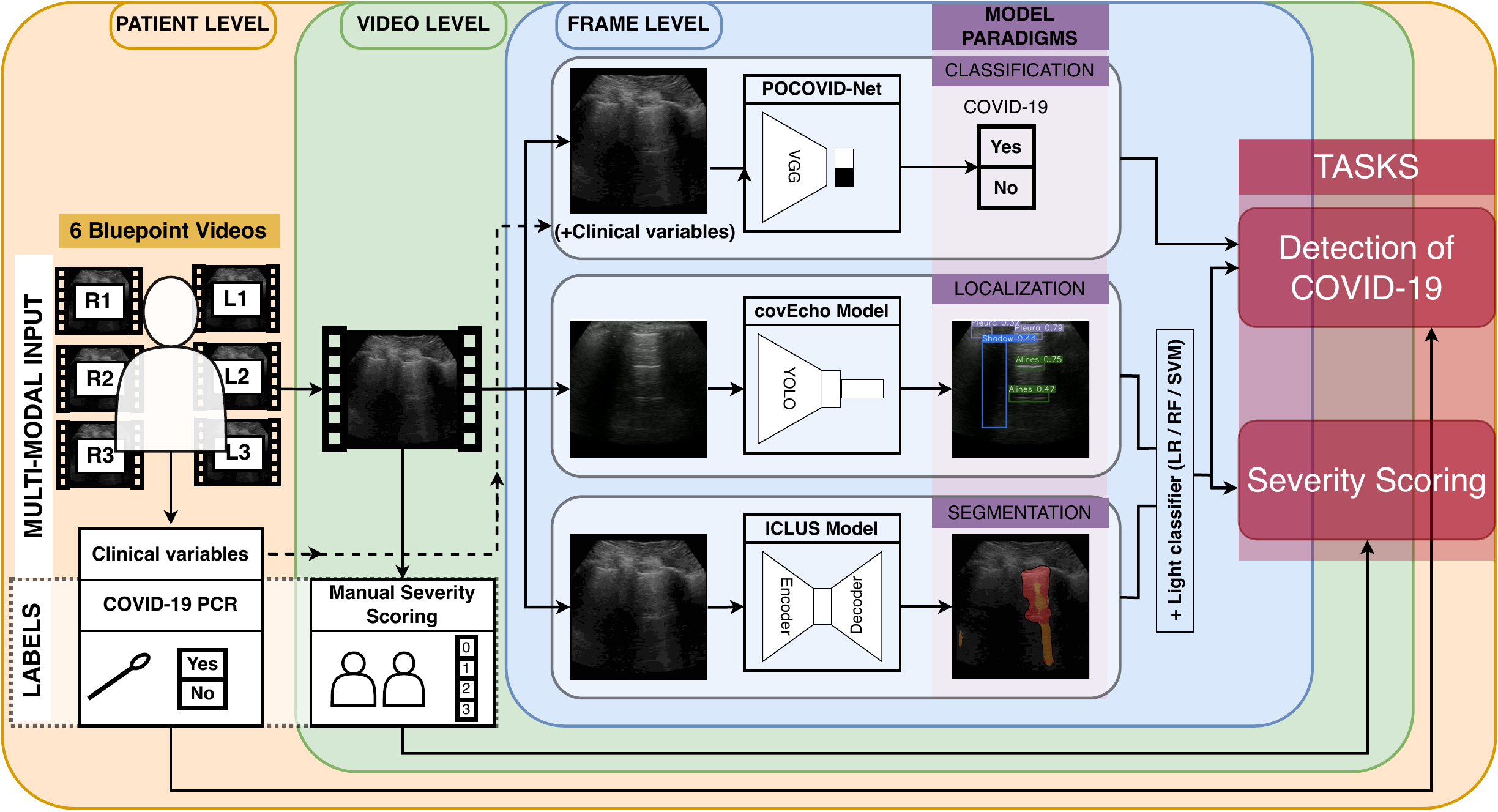}
    \caption{
    \emph{Overview:} In the COVID-BLUeS study, six LUS videos per patient (one per BLUE point) and clinical variables, including COVID-19 PCR results, were collected. The videos were labeled with severity scores by clinicians. The dataset is used to validate three seminal AI methods from the literature (classification, pattern localization and segmentation), focusing primarily on the tasks of COVID-19 detection and severity scoring. Localization and segmentation are paired with a light ML model for classification tasks. All three methods natively operate on the frame level. We evaluate the models on the video and patient level to investigate their value for clinical practice.
    }
    \label{fig:overview}
\end{figure*}

However, the validity of the results is generally limited by the representativeness of the used dataset. 
Datasets assembled from miscellaneous public sources suffer from high heterogeneity and biases (e.g., educational LUS videos are typically selected due to prominent lung involvement).
For LUS, this is particularly problematic, due to its diverse appearance dependent on the device, perspective and execution. 
Additionally, a large body of peer-reviewed AI models on LUS have reported unrealistically high performance in differential diagnosis~\citep{horry2020covid, al2021covid, saif2021capscovnet, karar2021lightweight, adedigba2021deep, sahoo2022covid,bruno2023efficient}, with some even reporting an accuracy of more than 98\%. 
Most likely, these artifacts are caused by meaningless splitting strategies between training and validation/test data; oftentimes the LUS videos were divided into 2D frames and then randomly split between train/test instead of patient-level split~\citep{al2021covid, saif2021capscovnet, karar2021lightweight, adedigba2021deep, sahoo2022covid,horry2020covid}. 
Overall, previously reported results only allow limited conclusions about the generalizability of the developed methods on new data and in clinical practice. 

Moreover, prospective studies on the performance of AI methods for lung ultrasound analysis are scarce~\cite{chen2021quantitative,nhat2023clinical} and none of them releases their data. 
Even though LUS is always only one out of many performed clinical examinations, it is largely unclear how LUS compares to, or can be combined with other clinical variables (e.g., anamnesis values or blood count) within AI models~\citep{mei-2020}. 
Last, existing work falls short of designated and fair comparisons between human experts and AI systems in assessing LUS data. 
LUS-based AI models are typically trained to either match human annotations (which are notoriously heterogeneous and suffer from low inter-operator agreement~\citep{lerchbaumer2021point}) or are evaluated against a non-US-based classification (e.g., PCR test).

What is the real value of the proposed AI methods for detecting and monitoring COVID-19? In this work, we set out to answer this question with a prospective study that assesses the performance of AI models in a controlled and realistic setting. 
We collect and fully release the largest public dataset of LUS for COVID-19, named COVID-BLUeS (BLUE point Lung Ultrasound). 
It contains $371$ videos from $63$ patients at Maastricht University Medical Center+ and a rich anamnesis including symptoms, blood count data, comorbidities, and clinical course. 
LUS acquisition followed the standardized BLUE protocol~\cite{lichtenstein2014lung} and only one US device type was used to ensure homogeneity. On this novel, highly standardized dataset, we systematically compare and combine seminal AI approaches for LUS analysis to assess their value among two clinical tasks: detection and severity scoring of COVID-19. 

Upon observing discrepancies between the results of the models on our dataset and their previously reported performances, we conducted in-depth experiments to uncover the root causes of these differences. 
Our analysis reveals critical limitations of previous work and extracts five novel insights to guide further model development that can be summarized into three main statements:

\begin{enumerate}[noitemsep,topsep=0pt,parsep=0pt,partopsep=0pt,leftmargin=15pt]
    \item Due to heterogeneous existing datasets, biased toward highly visible pathologies, the field has underestimated the difficulty of successfully applying AI on LUS data. 
    \item The video format of LUS data challenges frame-based classification models and leads to divergence between human severity assessment and AI-based pathology detections.
    \item Clinical variables and BLUE points were not considered before, due to a lack of data, despite their predictive power.
\end{enumerate}

\begin{figure*}[htb]
    \centering
    \begin{subfigure}[b]{0.235\textwidth}
    \includegraphics[width=\textwidth]{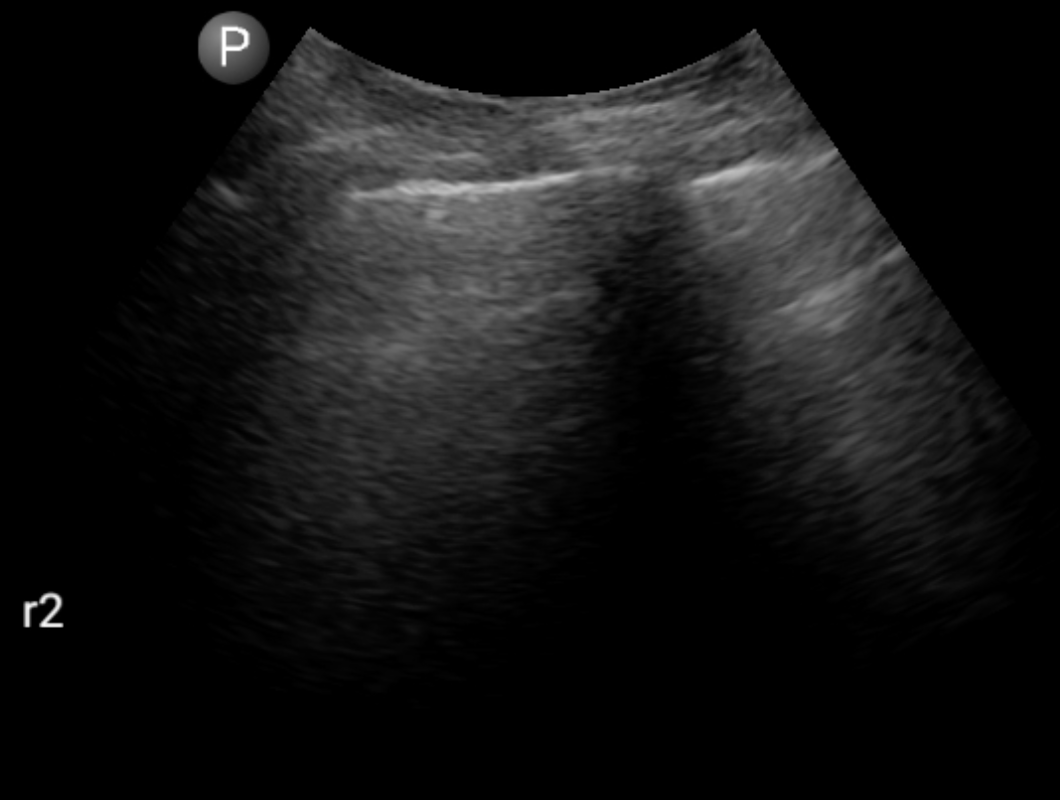}
    \caption{severity \textbf{0} (PCR neg.)}
    \label{fig:sev0}
    \end{subfigure}
    \begin{subfigure}[b]{0.235\textwidth}
    \includegraphics[width=\textwidth]{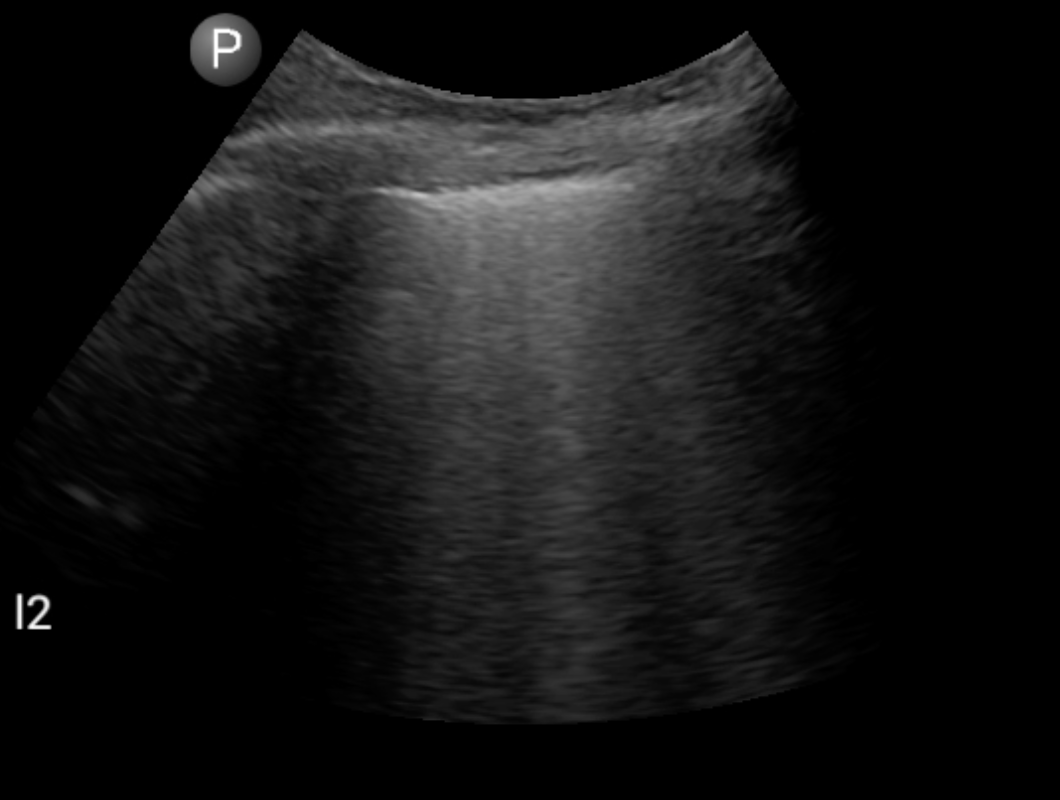}
    \caption{severity \textbf{1} (PCR neg.)}
    \label{fig:sev1}
    \end{subfigure}
    \begin{subfigure}[b]{0.235\textwidth}
    \includegraphics[width=\textwidth]{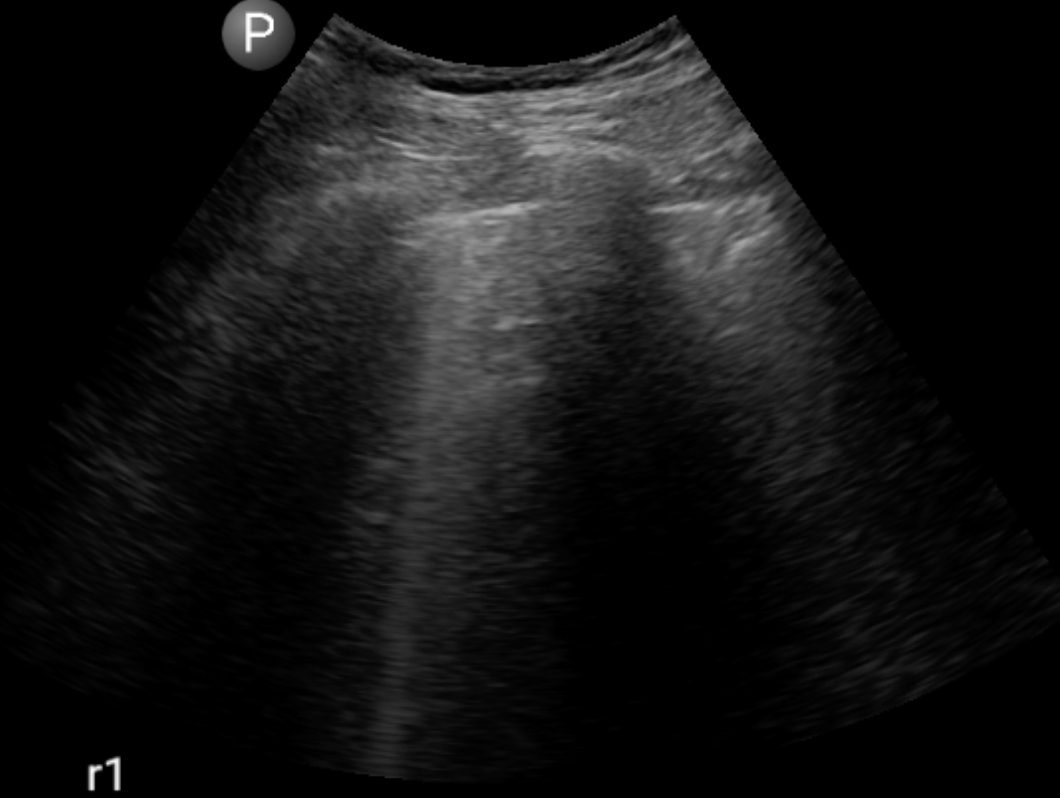}
    \caption{severity \textbf{2} (PCR pos.)}
    \label{fig:sev2}
    \end{subfigure}
    \begin{subfigure}[b]{0.235\textwidth}
    \includegraphics[width=\textwidth]{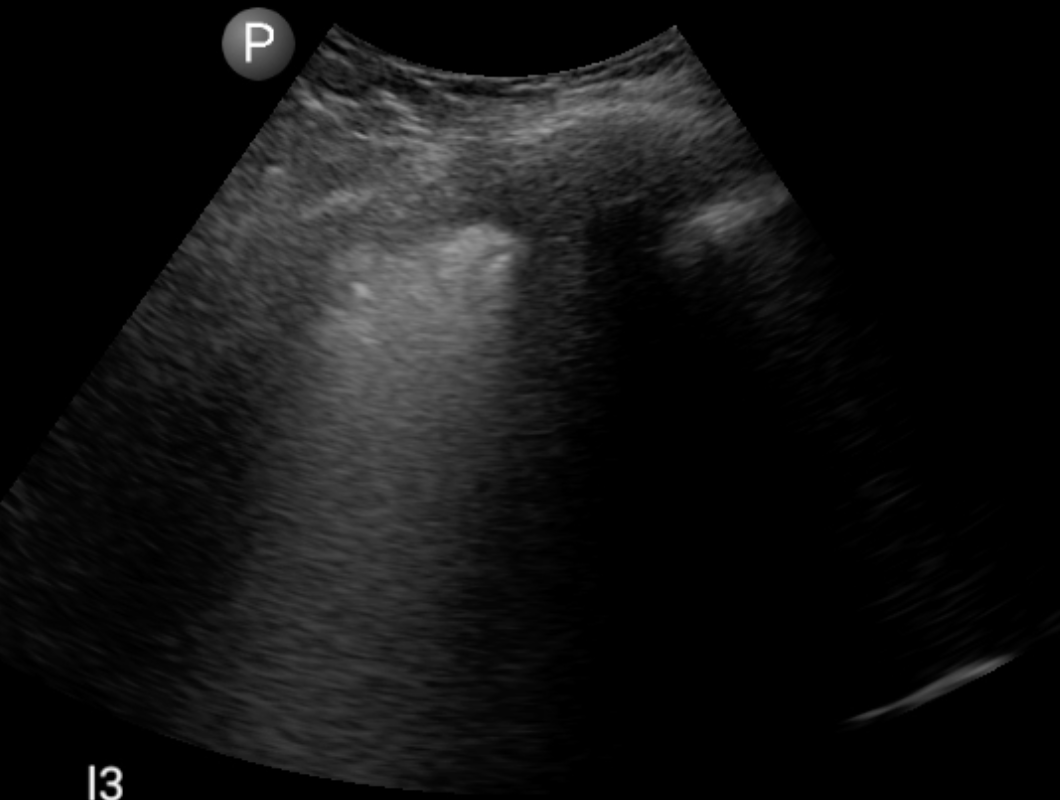}
    \caption{severity \textbf{3} (PCR pos.)}
    \label{fig:sev3}
    \end{subfigure}
    \caption{LUS example severity scores.
    (a) Severity 0: continuous, regular pleural line, A-lines are present.
    (b) Severity 1: Pleural line is indented. Mild vertical areas (B-lines) are visible.
    (c) Severity 2: Broken pleural line. Consolidated areas with patches of B-lines appear.
    (d) Severity 3: Dense and largely extended white lung with or without consolidations.
    }
    \label{fig:severity_examples}
\end{figure*}

\section{Methods and Materials}\label{sec:methods}

\subsection{Study design}\label{sec:data}

This prospective study was conducted at the emergency department of the Maastricht University Medical Center+ (MUMC+) and included all patients arriving in the emergency room that fulfilled inclusion criteria and agreed to study participation.
Inclusion criteria were self-reported COVID-19 symptoms (such as fever or chills, cough, shortness of breath or difficulty breathing, new loss of taste or smell, sore throat, congestion or runny nose), aged 18 years and above, and patients had to be capacitated (e.g. they were able to make a reasonable judgement of their own interests with regards to the study). Patients who were pregnant or had contra-indications for ultrasound were excluded. Eligible patients completed an informed consent, and afterwards, anamnesis, a complete blood count (CBC) and a PCR test for COVID-19 were conducted. In addition to routine care, six lung ultrasound videos (${\sim}3$ sec. each, frame rate of 20) were recorded by medical doctors experienced with US. Quality was ensured by adhering to the BLUE protocol~\citep{lichtenstein2014lung} and conducting a manual review of the videos to verify clear visibility of the lung. The same US device (Philips Lumify) with a convex probe\footnote{The LUS was conducted with a frequency between 4 and 5, no cosmetic filters, and by default a depth of 12cm and a gain of 36, but adjusted depending on the proband's body type. The focal area was set to the pleural line.} was used for all videos (see \autoref{fig:severity_examples}). 
The ultrasound videos were manually labeled for severity of lung involvement by two experts\footnote{4th year medical students with 3 years of experience as a student tutor for clinical ultrasound at BIKUS}, who split the videos. 
The severity was labeled on a scale from 0 to 3 according to the scoring algorithm by \citet{soldatiProposalInternationalStandardization2020}. 
It was noted whether A-lines and B-lines are visible. 
The experts were asked to judge solely based on the video, without any other information about the patient. The care provider, the US operator, and the labelers were blinded to the PCR test result.  
Examples for the videos with labeled severity are shown in \autoref{fig:severity_examples}.
Last, for consumption by ML models, the clinical variables obtained through anamnesis (cf.~\autoref{tab:variables_overview}) were standardized by removing the mean and scaling to unit variance.
Missing data\footnote{Missing data (number of missing patients): Blood pressure diastolic (1), duration of symptoms (2), Trombocytes (2), Leucocytes (1), Lymphocytes (7), LDH (3), pH (13), p02 (13), pCO2 (13), HC03 (13), Excess (13)} were imputed using the mean or median, based on a skewness test, i.e., for normally distributed variables, the median was used; otherwise, the mean was applied.

\subsection{Pre-trained machine learning models}
To assess the effectiveness of AI-based LUS analysis for detection and severity scoring of COVID-19, we employ three established ML models, representative for the tasks of classification, pattern detection, and segmentation (see \autoref{fig:overview}). Models were selected by performance, their prevail in the field, and by the availability of pre-trained models. 
All models operate on a \textit{frame}-level. 
Formally, we are given an image $x\in \mathbb{R}^{w\times h \times 3}$ and aim to learn a function $f_{c} : \mathbb{R}^{w\times h \times 3} \rightarrow \{0, 1\}$ to the COVID-19 classification, or to the severity score $f_{s}: \mathbb{R}^{w \times h \times 3} \rightarrow \{1, 2, \dots, C\}$ with $C$ classes. 

In addition to direct image-based classification, we consider two-step approaches, where pathological patterns are first segmented or localized on the image and later related to the diagnosis or the severity class. Segmentation yields one label per pixel $f_{seg}: \mathbb{R}^{w \times h \times 3} \rightarrow \mathbb{R}^{w \times h \times 3}$, whereas localization outputs bounding boxes $f_{l}: \mathbb{R}^{w \times h \times 3} \rightarrow \mathbb{R}^{d\times5}$, where $d$ is the number of localized pathologies, each represented by its class label and a bounding box (see \autoref{fig:overview}).

\subsubsection{Image classification via POCOVID-Net}

First, we approximate $f_{c}$ with the classification model \texttt{POCOVID-Net}~\cite{born2021accelerating} that leverages a Convolutional Neural Network (CNN), VGG-16, and produces a classification of the lung state into four classes (healthy,  infected with bacterial pneumonia, infected with viral pneumonia and uninformative~\cite{born2021accelerating}). 
The model passes the input image through 16 convolutional layers and a feed-forward head, and outputs a vector with softmax-probability for each of the four classes. Since the POCOVID dataset was assembled from various sources, it is prone to be very heterogeneous and potentially biased. To ensure compatibility with their pre-trained model, we replicate their data preprocessing steps, specifically cropping the videos to a size of 224 x 224 and normalizing pixel values to the [0,1] range.

\subsubsection{Pattern localization and severity scoring with the CovEcho model}\label{sec:covecho}

The CovEcho model~\citep{josephCovEchoResourceConstrained2022} deploys a YOLOv5~\citep{redmonYouOnlyLook2016} object detection model trained on a proprietary dataset of 570 annotated LUS images from both linear and convex probes.  The model is trained to detect important LUS landmarks ($f_{l}$) and the pre-trained model was made available to us. YOLOv5 is a fast and efficient single-stage object detection architecture. It predicts bounding box locations and class probabilities directly from the input image (see \autoref{fig:overview}) by dividing it into a grid, where each cell predicts bounding boxes, confidence scores, and labels for objects within its scope. This streamlined approach contrasts with two-stage detectors, which separate the region proposal and classification stages, making YOLO significantly faster while maintaining high accuracy. 
Notably, the CovEcho model assigns a quality score between 0 and 100 to each frame, dependent on the detected landmarks (see Appendix~B). 
Of the $\sim$32k frames from our dataset, $\sim$18k were excluded from the analysis since their quality score was below 50. On average the model discarded $46.5$ ($\pm33.5$) frames per video, for example where the pleura could not be detected. 

\subsubsection{Pattern segmentation with the ICLUS model}\label{sec:iclus}

\citet{roy2020deep} proposed an ensemble of three CNN models for segmentation (approximating $f_{seg}$), namely U-Net~\citep{ronnebergerUNetConvolutionalNetworks2015}, its successor U-Net++~\citep{zhouUNetNestedUNet2018}, and DeepLabV3+~\citep{chenEncoderDecoderAtrousSeparable2018}. All three models are popular encoder-decoder architectures with many residual connections between encoder and decoder to ensure information transfer. 
\citet{roy2020deep} trained these models separately on the ICLUS dataset of 1005 manually labeled LUS frames. In contrast to the CovEcho model, it produces a pixel-wise segmentation of the image, as shown in \autoref{fig:overview}. They report a Dice coefficient of 0.75 for their final ensemble model that aggregates the three predictions. 

\subsection{Classification based on pattern-based models}\label{app:lrrfsvm}

For comparing classification to segmentation or localization models for the task of COVID-19 detection, we implement a two-stage approach to predict the PCR test result from the outputs of the CovEcho and ICLUS models (see \autoref{fig:overview}). 
First, the CovEcho or ICLUS model is applied to segment or detect pathological patterns. In the second step, features of these detected patterns, specifically the class count and class area for the CovEcho model and the class-wise pixel-count area of the ICLUS model (see \autoref{fig:classes_severity}), are provided to a simple ML model that is trained for COVID-19 classification. Specifically, we tested a Logistic Regression (LR) model, a Random Forest (RF) Classifier and a Support Vector Machine (SVM) Classifier for the second stage. Formally, the localization model $f_l$ (cf. \ref{sec:covecho}) outputs bounding boxes $\in \mathbb{R}^{d\times 5}$ and is combined with a classifier $g_{l}: \mathbb{R}^{d\times 5} \rightarrow \{0, 1\}$, such that the PCR test result is predicted from the raw image as $g_{l} \circ f_{l}$. Correspondingly, the segmentation model $f_{seq}$ (cf. \ref{sec:iclus}) is combined with $g_{seg}: \mathbb{R}^{w \times h \times 3} \rightarrow \{0, 1\}$ and applied as $g_{seg} \circ f_{seg}$.

\subsection{Training multi-modal CNN models}

Since dataset differences may prevent the successful application of \textit{pre-trained} models to our dataset, we retrain the model from \citet{born2021accelerating} on the COVID-BLUeS dataset, alongside other commonly used CNNs. For training details, see Appendix D. In contrast to previous work that was predominantly bound to image datasets without patient data, we further test a multi-modal approach combining images and clinical variables.
In this case, the clinical variables from \autoref{tab:variables_overview} are encoded in a vector (binary variables are one-hot-encoded), and are passed through one fully-connected layer before being concatenated with the output of the last convolutional layer. This combined embedding of image and clinical variables is passed through three dense layers to output a classification.

\subsection{Statistical analysis}

To evaluate the ability of AI methods to detect COVID-19, we compute the accuracy ($\frac{TP+TN}{TP+TN+FN+FP}$), F-score ($\frac{2TP}{2TP + FP + FN}$) the sensitivity ($\frac{TP}{TP + FN}$) and the specificity ($\frac{TN}{TN + FP}$). The correspondence between the severity scores and predicted severity is assessed via a Spearman rank correlation test, and as a multi-class classification problem (4 severity labels) via accuracy, average class-wise sensitivity, and average $F_1$-score ($\frac{2TP}{2TP+FP+FN}$). 
Note that these metrics can be computed on a frame, video or patient level (see \autoref{fig:overview}), as the employed ML methods provide frame-wise prediction. 
If not denoted otherwise, the results are reported on a \textit{patient} level, where the frame-wise outputs are averaged per patient. 
For finetuned models, we report the results from 5-fold cross validation; i.e., splitting the data into five parts and performing five experiments where training is done on four parts and testing on the fifth part.
The data is split by \emph{patients}; i.e., the videos of one patient are never in both train and test set. 
All models are trained on a frame-level, hence they are exposed to ca. 25K training samples. Data augmentation was used to increase the sample size (see Appendix D). 

\section{Dataset and study population}
\label{sec:studypop}
Between February and May 2021, 68 patients were enrolled in the study. 
While for 43 patients, one video per BLUE point was obtained as desired, for 10 patients there are more than six videos due to repeated examination. 
Five patients were excluded entirely since the LUS data was missing. 
Vaccination data was scarcely available and is thus not further analyzed.
After the severity scoring process, eight videos from five patients were excluded since the lung was not visible, leaving 371 videos with 31746 frames in total.
The most dominant SARS-CoV-2 mutation was the Alpha variant. Based on PCR results, 33 patients (52\%) were positive for COVID-19. Only one patient was diagnosed with other viral pneumonia and 3 with bacterial pneumonia, whereas 16 patients had other lung diseases and for the remaining 10 patients the lung was healthy.  \autoref{tab:variables_overview} shows baseline characteristics for the total cohort, and split by COVID-19 diagnosis. Significant differences according to a Chi-Square test for binary and a Mann-Whitney U test for continuous variables are marked. \autoref{tab:variables_overview} shows that the blood parameters of LDH, CRP and HG are positively correlated and the oxygen saturation is negatively correlated with the PCR test result.
The analysis reveals that the severity scores assigned by human experts are not significantly different between COVID-19 positive and negative patients ($p=0.89$, $t$; Spearman $\rho = 0.007$); i.e., patients with a COVID-19 infection are not more likely to show serious symptoms in the lung. 
Further results on severity scores are shown in Appendix~A. 

\input{figures/table_clinical_variables}

\section{Results and insights}
\label{sec:results1}
The key results for AI-based COVID-19 detection, evaluated against PCR test outcomes, are summarized in \autoref{tab:CNN_perform}. 
The human diagnosis was obtained by classifying patients with B-lines present in four or more videos as positive. Other rule-based classification schemes, including those incorporating severity scores, consistently showed lower accuracies (see Appendix H).  
Notably, human annotators achieved a maximum accuracy of 65\% at the patient level, significantly underperforming compared to the AI models. The best performance was achieved using the ICLUS segmentation model combined with logistic regression, yielding an accuracy of 79\%. However, even this model falls significantly short of the higher accuracies reported in related work. 
These findings highlight the substantial challenges in accurately classifying the COVID-BLUeS dataset, regardless of the method—whether through human assessment, zero-shot application of AI models, CNN training, or pattern detection techniques. 
In the following, we therefore present an in-depth analysis of these challenges, highlighting the discrepancies with and limitations of prior studies. Our analysis is structured into five key statements addressing the difficulties of applying AI methods in clinical practice.

\subsection{
Performance claimed in previous work does not reflect the complexity of realistic LUS datasets
}


Ideally, ML models should be applicable to unseen LUS data, without requiring calibration for a new clinical setting. 
Therefore, we first evaluate the model outputs when applied directly on the new data (``zero-shot'').
We validate the model by \citet{born2021accelerating}, who reported an accuracy of 87.8\% on the POCUS dataset~\citep{born2021accelerating}, in a zero-shot setting on the COVID-BLUeS dataset. 
The model achieved only an accuracy of 65\% (sensitivity 0.62, specificity 0.70) when the frame-wise outputs are aggregated per patient. 
On the positive side, even though the model was originally trained to detect \textit{four} classes (healthy, viral pneumonia, bacterial pneumonia and uninformative), all samples were either classified as healthy or viral pneumonia, demonstrating an ability to exclude a diagnosis of bacterial pneumonia. 
The difference with respect to the previously reported model performance may be explained through the heterogeneity of the POCUS dataset, in particular the inclusion of educational videos, biasing it towards samples with clearly visible biomarkers.
%
Notably, POCOVID-Net served as an example here but this issue extends to the  majority of prior art the field. Many studies either (1) train  on the POCOVID dataset~\citep{Al-Jumaili-2021, karar2021lightweight, robertsUltrasoundDiagnosisCOVID192020, saif_capscovnet_2021, macleanCOVIDNetUSTailored2021, adedigba2021deep, bruno2023efficient}, (2) rely on manually selected frames or clips that clearly exhibit pathologies~\citep{la_salvia_deep_2021, arntfieldDevelopmentConvolutionalNeural2021}, or (3) use image datasets requiring clinicians to pause and save ultrasound examinations at suitable moments~\citep{chen2021quantitative}. 
Such existing datasets are biased towards manual selections thus hindering zero-shot application on unbiased videos. 
Beyond dataset biases, the reproducibility of certain 
peer-reviewed AI models on LUS is highly questionable as they have reported unrealistically high performance in differential diagnosis, surpassing $99\%$~\citep{al2021covid,saif2021capscovnet,karar2021lightweight}, sometimes even $100\%$~\cite{horry2020covid,bruno2023efficient} -- likely due to flawed data splitting.

\subsection{
Specialized classification (or diagnosis) models can be outperformed with generalist segmentation models
}

When training various popular CNN-classifiers, we find that, in line with~\citet{born2021accelerating}, the classification performance is highest for POCOVID-Net (cf. \autoref{tab:CNN_perform}, block 2, technical details in Appendix D). 
\input{figures/table_main_results}
However, the accuracy peaks at 73.4\% despite parameter tuning, indicating that frame-based classification may indeed be significantly harder than on previous datasets. 
In comparison, we test the two-step approach (cf. Section~\ref{app:lrrfsvm}) using the
CovEcho localization and the ICLUS segmentation models together with a light classifier (LR / RF / SVM). 
As shown in \autoref{tab:CNN_perform} (block 3), the CovEcho-based models achieved an accuracy of up to 66\% (patient-level) in predicting the PCR test result, whereas the ICLUS model yields an accuracy of 79\%.  
Prediction via logistic regression on the segmented class area is thus the best image-based model for diagnostic purposes in our experiments, outperforming frame-wise classification. This is remarkable as only the light classifier is trained on the new dataset, while the base model operates zero-shot, unlike the fully trained CNN models. 
The better accuracy of ICLUS-inputs compared to CovEcho-inputs is due to the higher information content of segmentations, in contrast to the simple class count of the CovEcho model. 
The superiority of segmentation- over classification models may be explained by uninformative frames distracting the classification result. 
This is less problematic for segmentation models since their classification is based on the total observation of pathologies across the video. 

\subsection{Clinical variables alone present a strong baseline that is missing in related work.}\label{sec:clinical}

To assess the predictive power for the PCR test outcome, we distinguish three scenarios to combine LUS videos, clinical variables and blood count.
First, we only utilize the LUS images.
Secondly, we combine LUS with 55 clinical variables.
Thirdly, we also add 11 CBC variables (see \autoref{tab:variables_overview} and Appendix D). 
To the best of our knowledge, this is the first analysis combining and comparing LUS images with clinical variables.

\input{figures/table_variables_results}
%
As a baseline, we set a Random Forest (RF) model \textit{that does not use images} and achieves $64\%$ accuracy only with clinical variables and $84\%$ with clinical and CBC data. 
This confirmed previous claims that blood test data is indicative for COVID-19 infections~\citep{aljame2020ensemble}, the RF primarily relied on CRP, LDH and HG values (see Appendix C). 

The results are shown in \autoref{tab:CNN_perform_variables}. 
The best multi-modal model achieves 83\% accuracy (ResNet50 with LUS + Clinical variables), clearly outperforming the best image-only model (POCOVID-Net; 73.4\%). 
Critically, however, it does \textit{not} outperform the RF on tabular features.
This is an important finding which highlights the importance to include non-image baselines when studying the value of LUS-based models. 
We found this missing in all related work.

\subsection{AI-detected pathological patterns are only weakly correlated to human-based severity assessment
}
Severity scoring as another major task in LUS analysis (cf. \autoref{fig:overview}). 
The CovEcho model~\cite{josephCovEchoResourceConstrained2022} localizes landmarks such as A-lines, B-lines and consolidations on LUS images via bounding-boxes, whereas the ICLUS model~\cite{roy2020deep} provides a direct measure of the area of the detected landmark by providing a precise boundary via segmentation. 
We first analyze the correlation of the detected pathologies to the human-annotated severity scores and then quantify their predictive value. 

\subsubsection{Analytical results}

\begin{figure*}
    \centering
    \includegraphics[width=\textwidth]{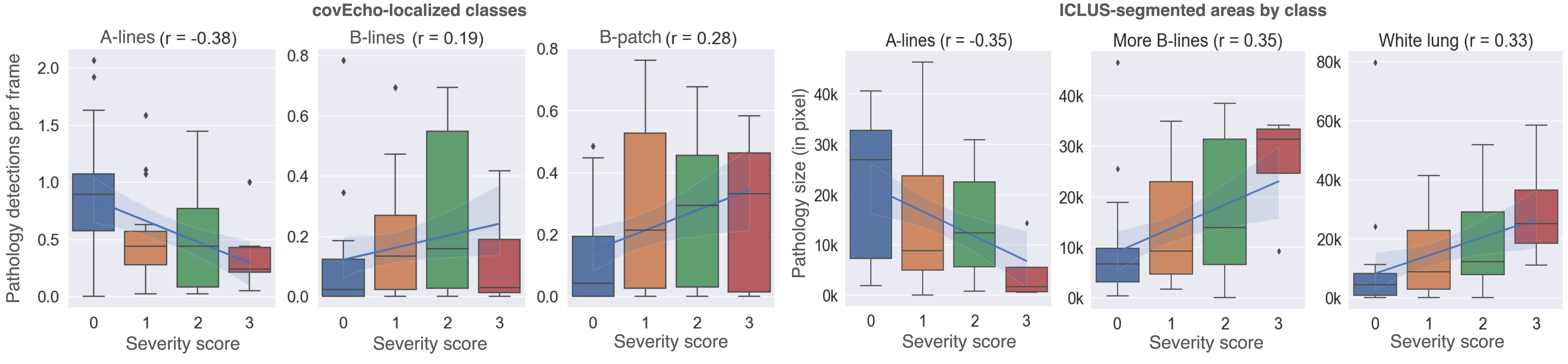}
    \caption{
    Number of detected pathologies (CovEcho / ICLUS models) by severity score (only selected pathologies are shown). As expected, typical signs for a COVID-19 detection, such as B-lines, are detected more frequently (higher number of pathology detections / size) for samples with higher severity score. Conversely, patterns indicating a healthy lung, such as A-lines, are observed more often in samples with low severity scores. Pearson r correlations are provided above each plot and as regression lines.}
    \label{fig:classes_severity}
\end{figure*}

In \autoref{fig:classes_severity} we relate the count and the area of detected pathologies to the human severity score. 
The number of A-lines detected by the CovEcho model correlates negatively with the severity score.
Next, a higher number of detected B-lines and B-patches is related to a higher severity. 
A similar pattern is found for the segmented area of the ICLUS model; in particular, the class ``More B-lines'' clearly correlates with the severity score ($r=0.35$). Further analysis in Appendix G. 
shows significant differences between the pixel count distributions of positive and negative patients.

\citet{josephCovEchoResourceConstrained2022} further proSpose a scheme to derive severity scores from the patterns detected with the CovEcho model. In Appendix B 
we show that the derived severity scores correlate weakly with the severity 
labels ($\rho=0.1$). 
The CovEcho severity scoring seems to be  more conservative than the labelers of our study, assigning lower scores on average.
\subsubsection{Predictive performance of severity scoring}\label{sec:results2_3}
We trained basic ML models such as LR, RF and SVM to predict the severity score from the detected class count and area. The results of the best models are given in \autoref{tab:severity_score_prediction_covEcho} (see Appendix E 
for confusion matrices), testifying a maximal accuracy of 54\% in severity classification. The mean absolute error (MAE) between the predicted score and the real score is 1.02 for the CovEcho model and 1.07 for the ICLUS-segmentation model (RMSE of 1.36 and 1.41 respectively). 
While humans judge the severity typically based on a few informative frames, the ML models perform a post-processing aggregation of \textit{all} frame-wise results. 
Our attempts to improve the aggregation, e.g., by selecting only frames with high confidence score, failed to yield clear improvements.

\begin{table}[h!t]
    \centering
    \resizebox{0.85\linewidth}{!}{
    \begin{tabular}{lrrr}
        \toprule
        & \multicolumn{3}{c}{CovEcho severity classification}\\
     & Accuracy & Precision & F1 score \\
        Level &  &  &  \\
        \midrule
        Video  &  0.33 $_{\pm 0.05}$ &  0.40 $_{\pm 0.09}$ &  0.32 $_{\pm 0.04}$  \\
        Patient &  0.54 $_{\pm 0.12}$ &  0.52 $_{\pm 0.10}$ &  0.51 $_{\pm 0.11}$ \\
    \bottomrule\\
    \end{tabular}
    }
    \resizebox{0.85\linewidth}{!}{
    \begin{tabular}{lrrr}
        \toprule
        & \multicolumn{3}{c}{ICLUS-segmentation severity classification} \\
        & Accuracy & Precision & F1 score \\
        Level &  &  &  \\
        \midrule
        Video  &  0.40 $_{\pm 0.06}$ &  0.40 $_{\pm 0.05}$  &  0.39 $_{\pm 0.06}$ \\
        Patient  &  0.51 $_{\pm 0.09}$ &  0.46 $_{\pm 0.11}$ &  0.48 $_{\pm 0.09}$ \\
    \bottomrule\\
    \end{tabular}
    }
    \caption{Video and patient-level results for the prediction of the labeled severity score, obtained with CovEcho and ICLUS models together with a light classifier (RF / SVM / LR). Only the best performing combinations are shown.
    }
    \label{tab:severity_score_prediction_covEcho}
\end{table}

\subsection{
LUS of the anterior lung has higher predictive power than the lateral or posterior lung }

\begin{figure*}[htb]
    \centering
    \includegraphics[width=0.9\textwidth]{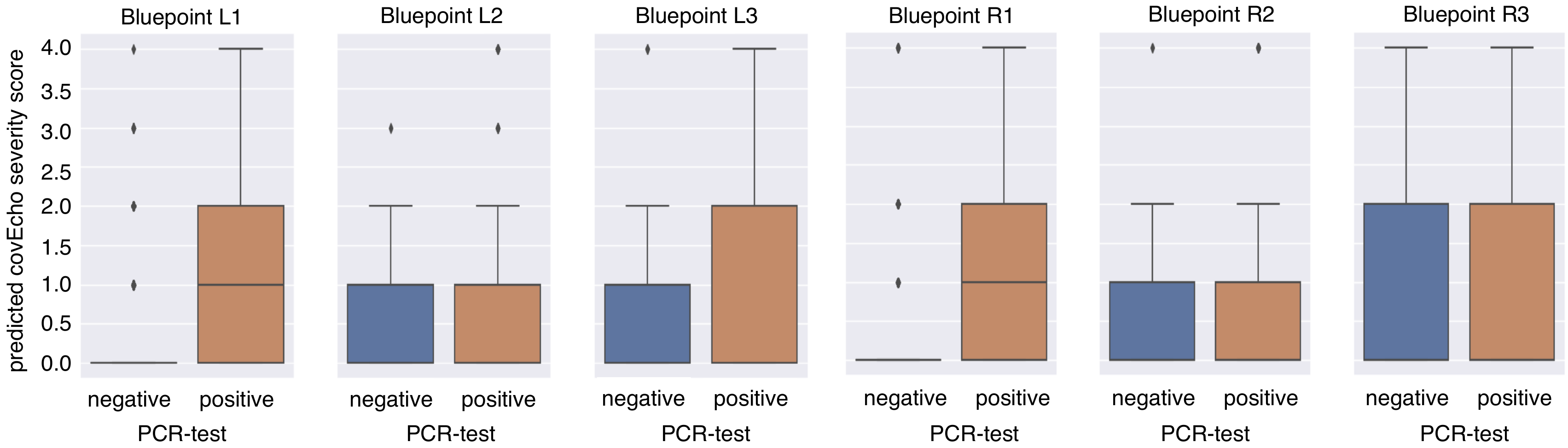}
    \caption{Plotting the severity scores as predicted by the CovEcho model, separated by the COVID-19 diagnosis for each BLUE point. This comparison clearly suggests that the model performance differs significantly between BLUE points. The plot also shows a symmetry in the results between the left and right side affirming that the results are not coincidental.
    }
    \label{fig:bluepoints}
\end{figure*}

To the best of our knowledge, this is the first study on AI-based US classification that distinguishes between BLUE points.
The severity scores predicted with the CovEcho model strongly depend on the BLUE point. 
As \autoref{fig:bluepoints} demonstrates, the relation between a COVID-19 infection (PCR test positive / negative) and the CovEcho severity score is much more pronounced for the anterior BLUE points (L1 and R1) compared to lateral and posterior BLUE points.

\section{Discussion}\label{sec:discussion}

Our study identified general difficulties in the application of ML models to data collected in a realistic clinical setting:
First, we found that zero-shot application of existing ML models to predict the PCR test result yields low accuracy ($\sim 65\%$).
Secondly, retraining the models on the COVID-BLUeS data provided significant improvements, but could not match reported performance of more than 90\% accuracy in COVID-19 diagnosis~\citep{al2021covid, saif2021capscovnet, karar2021lightweight, adedigba2021deep, sahoo2022covid,bruno2023efficient,josephCovEchoResourceConstrained2022}, even when using the same models.
Thirdly, we could also not match the severity prediction performance of 0.71 ($F_1$-score) as reported on the ICLUS data~\citep{roy2020deep}. 
But it has to be emphasized that human experts had enormous difficulties classifying our dataset solely based on the LUS videos. 
Surprisingly, expert-assigned COVID-19 severity scores did not differ between COVID-19-positive and negative patients. 
Thus, when looking beyond comparisons to existing datasets and seeking for an internal comparison, we found that the best ML models still outperform the human experts in all settings and metrics by a large margin.

Compared to existing LUS datasets~\cite{born2021accelerating,roy2020deep,josephCovEchoResourceConstrained2022}, classifying our COVID-BLUeS dataset is more challenging due to the heterogeneity of the study population, the lack of recording biases (e.g., device heterogeneity) and inclusion biases, a lack of preprocessing to identify informative frames and its prospective nature. 
On top of that, some results in previous work are  erroneous due to a video-agnostic train-test split (i.e., images of one video are partly used as training data). 
Indeed, even on the COVID-BLUeS dataset we find an accuracy of 98\% with such a train-test split.

Our findings are in line with previously reported high predictive power of CBC-based models~\citep{aljame2020ensemble}. 
Notably, the COVID-BLUeS dataset allows for a more fine-grained analysis than previous studies on AI for LUS because it differentiates among BLUE points. 
Our analysis showed striking differences in the results for different parts of the lung, indicating higher predictive power at BLUE points L1 and R1, in contrast to the findings of \citet{perrone2021new} reporting higher severity in lower parts of the lung.

Aside from the need to validate new ML methodology for LUS on representative datasets, our findings thus call for an increased awareness in ML research to the unique challenges involved with US imaging, such as operator dependency, data heterogeneity and its temporal nature. 
Contemporary AI models have not considered the spatial location of a LUS scan which could be a future promising endeavor, e.g., to build a temporal 3D-model of lung involvement. Our study introduced several novel directions for LUS model development, such as consideration of BLUE points or the integration of clinical variables. Additionally, it provides guidelines for selecting among existing approaches. When sufficient data is available for training or fine-tuning, CNN-based image classification remains most effective, at least when combined with clinical variables. However, if data is limited, leveraging pre-trained segmentation models with a lightweight classifier is preferable. Regardless of the chosen approach, it is crucial to ensure that the data distribution aligns with the training data when applying pre-trained models.

From a clinical perspective, our study implies that AI methods are relevant as decision-support tools in triage, but are not sufficiently advanced to overrule human assessment in any way. Interdisciplinary collaborations are necessary to align the skills of ML engineers with the needs of clinical practitioners and to properly assess the potential of AI in LUS analysis. Importantly, the results of this study are relevant for AI-based diagnosis of other lung conditions. COVID-19 is a suitable test case due its large prevalence and the abundance of AI approaches; however, our findings provide guidelines on general AI-based LUS analysis. This is particularly relevant for diseases where PCR or antigen testing are not available.

The study is limited by the size and heterogeneity of the study population, and by the selection of three AI models (not including video-based analysis), but an extensive comparison of models was not the purpose of our study. Instead, we aimed to gain a better understanding of the difficulties involved with lung ultrasound analysis and to assess the real-world value of published methods. With the publication of the high-quality COVID-BLUeS dataset, we hope to inspire further research that bridges the gap between ML modelling and clinical practice. 

\section*{Acknowledgments}
We deeply thank Prof. Ender Konukoglu for providing access to the data storage and compute infrastructure of D-ITET, ETH Zurich which was fundamental for successful execution of the project.
We also thank Prof. Bastian Rieck for useful discussions about topological data analysis. 
We thank Prof. Ruud van Sloun for giving access to the trained ICLUS model.

\section*{Data availability}

The dataset is available on GitHub under a CC-NC-ND license: \url{https://github.com/NinaWie/COVID-BLUES}.

\section*{Patient consent}
All subjects gave their written informed consent, and the study protocol was reviewed and approved by the institute’s committee on human research to be exempted from the Dutch Medical Research with Human Subjects Law: METC azM Maastricht, number [METC 2020-2229].

\section*{Funding}
This research received no external funding.

\section*{Author contributions statement}

\noindent Conceptualization: JB, NW, DdKdB \\
Data Curation: NW, CB, FAME, SR, SA, CMER \\ 
Formal Analysis: NW, MR, SR, CB, FAME \\
Funding Acquisition: WB \\ 
Investigation: RH, SA, SvdW, CB, FAME, DdKdB, LG  \\ 
Methodology: JB, NW, MR, SR, DdKdB\\
Project Administration: DdKdB, RH, SvdW, LG  \\
Resources: SvdW, EK, RH, WB \\
Software: JB, NW, MR, SR\\
Supervision: JB, NW, DdKdB, WB \\
Validation: JB, NW, MR\\
Visualization: NW, MR \\
Writing -- Original Draft: JB, NW, MR\\
Writing -- Review: JB, NW, SvdW, CB, DdKdB, RH, WB \\




\FloatBarrier
\bibliographystyle{plainnat}
\bibliography{references}

\input{appendix}

\end{document}

%% file: figures/table_clinical_variables.tex
\begin{table*}
\resizebox{\textwidth}{!}{
\begin{tabular}{lrrrr}
\toprule
            &         
            \begin{tabular}{c} \textbf{Total} \\ \textbf{(n=63)} \end{tabular}
            & \begin{tabular}{c} \textbf{COVID-19-positive}  \\ \textbf{(n=33)} \end{tabular} 
           & \begin{tabular}{c} \textbf{COVID-19-negative } \\ \textbf{(n=30)} \end{tabular}
           & \begin{tabular}{c} \textbf{p-value}  \\ \textbf{(group difference)} \end{tabular} \\
\midrule \multicolumn{5}{l}{\textbf{Patient characteristics}} \\ \midrule 
Age (years)                          &  
65.2 ($\pm$ 15.7)
&    66.2 ($\pm$ 15.7)
&     64.1 ($\pm$ 15.8)
&                         0.33 \\
Women                                &             41 (65\%) &                19 (57\%) &                22 (73\%) &                         0.30 \\
Height (cm)                          &    171.7 ($\pm$ 9.4) &       169.8 ($\pm$ 9.3) &       173.7 ($\pm$ 9.1) &                         0.07 \\
Weight (kg)                          &     81.3 ($\pm$ 19.4) &       81.2 ($\pm$ 16.3) &       81.5 ($\pm$ 22.3) &                         0.45 \\
BMI (kg/m2)                          &     27.7 ($\pm$ 6.3) &        28.3 ($\pm$ 5.1) &        27.0 ($\pm$ 7.3) &                         0.12 \\
Smoking                              &             29 (46\%) &                 9 (27\%) &                20 (66\%) &                        < 0.01(*) \\
\midrule \multicolumn{5}{l}{\textbf{Symptoms}} \\ \midrule 
Fever                                &             21 (33\%) &                13 (39\%) &                 8 (26\%) &                         0.42 \\
Cough                                &             40 (63\%) &                24 (72\%) &                16 (53\%) &                         0.18 \\
Sore throat                          &               3 (4\%) &                  2 (6\%) &                  1 (3\%) &                         0.93 \\
Running nose                         &               4 (6\%) &                  1 (3\%) &                 3 (10\%) &                         0.54 \\
Chest pain                           &             15 (23\%) &                 5 (15\%) &                10 (33\%) &                         0.16 \\
Muscle pain                          &               6 (9\%) &                 6 (18\%) &                  0 (0\%) &                        < 0.01 (*)  \\
Fatigue                              &             13 (20\%) &                10 (30\%) &                 3 (10\%) &                         0.09 \\
Headache                             &              7 (11\%) &                 5 (15\%) &                  2 (6\%) &                         0.50 \\
Number of applying symptoms (max. 8) &      1.7 ($\pm$ 1.3) &         2.1 ($\pm$ 1.4) &         1.2 ($\pm$ 0.9) &                        < 0.01 (*)  \\
Duration of symptoms (days)          &    10.6 ($\pm$ 26.5) &         8.1 ($\pm$ 5.1) &       13.4 ($\pm$ 37.9) &                        < 0.01 (*)  \\
\midrule \multicolumn{5}{l}{\textbf{Vital parameters}} \\ \midrule 
Oxygen saturation (\%)               &     92.6 ($\pm$ 5.0) &        90.6 ($\pm$ 4.9) &        94.7 ($\pm$ 4.0) &                       < 0.01 (*) \\
Respiratory rate (/min)              &     21.0 ($\pm$ 5.4) &         22.5 ($\pm$ 4.7) &        19.3 ($\pm$ 5.6) &                        < 0.01 (*) \\
Pulse (bpm)                          &    92.6 ($\pm$ 21.7) &       95.1 ($\pm$ 21.1) &        89.8 ($\pm$ 21.9) &                         0.14 \\
Body temperature (°C)                &     37.3 ($\pm$ 0.9) &        37.5 ($\pm$ 0.8) &         37.0 ($\pm$ 0.9) &                        < 0.01 (*) \\
Blood pressure systolic (mmHg)       &   130.7 ($\pm$ 20.2) &      132.7 ($\pm$ 20.3) &      128.5 ($\pm$ 19.7) &                         0.27 \\
Blood pressure diastolic (mmHg)      &    78.2 ($\pm$ 14.5) &       79.1 ($\pm$ 11.5) &       77.2 ($\pm$ 17.0) &                         0.42 \\

\midrule \multicolumn{5}{l}{\textbf{History of pulmonary diseases}} \\ \midrule 
Any pulmonary disease                &             36 (57\%) &                19 (57\%) &                17 (56\%) &                         0.86 \\
Asthma                               &               5 (7\%) &                 4 (12\%) &                  1 (3\%) &                         0.41 \\
COPD                                 &             13 (20\%) &                 6 (18\%) &                 7 (23\%) &                         0.85 \\
Pneunomia                            &              7 (11\%) &                 4 (12\%) &                 3 (10\%) &                         0.89 \\
Interstitial Lung Disease            &               1 (1\%) &                  1 (3\%) &                  0 (0\%) &                         0.96 \\
\midrule \multicolumn{5}{l}{\textbf{Blood parameters}} \\ \midrule 
pH                                   &      7.5 ($\pm$ 0.1) &         7.5 ($\pm$ 0.1) &         7.5 ($\pm$ 0.1) &                        0.04 (*) \\
pCO2 (kPa)                           &       4.4 ($\pm$ 1.0) &         4.3 ($\pm$ 0.5) &         4.6 ($\pm$ 1.3) &                         0.14 \\
p02 (kPa)                            &      9.5 ($\pm$ 3.0) &         9.3 ($\pm$ 2.0) &         9.7 ($\pm$ 3.8) &                         0.44 \\
HC03 (mmol/l)                        &     23.5 ($\pm$ 4.7) &        23.1 ($\pm$ 3.2) &        24.1 ($\pm$ 6.1) &                         0.27 \\
Excess (mmol/l)                      &      0.3 ($\pm$ 4.2) &         0.2 ($\pm$ 3.1) &         0.6 ($\pm$ 5.2) &                         0.45 \\
HG (mmol/l)                          &      8.0 ($\pm$ 1.4) &         8.4 ($\pm$ 0.9) &         7.5 ($\pm$ 1.6) &                        < 0.01 (*) \\
Trombocytes ($10^9/l$)               &  232.8 ($\pm$ 112.0) &     217.6 ($\pm$ 100.1) &     248.0 ($\pm$ 120.9) &                         0.14 \\
Leucocytes ($10^9/l$)                &      9.4 ($\pm$ 9.9) &         7.4 ($\pm$ 3.9) &       11.5 ($\pm$ 13.4) &                        0.03 (*) \\
Lymphocytes ($10^9/l$)               &    17.5 ($\pm$ 15.5) &       14.1 ($\pm$ 10.3) &       21.3 ($\pm$ 19.2) &                         0.06 \\
LDH (U/l)                            &  311.8 ($\pm$ 129.3) &     379.2 ($\pm$ 133.9) &      239.8 ($\pm$ 73.3) &                        < 0.01 (*)  \\
CRP (mg/l)                           &    82.2 ($\pm$ 82.6) &      110.4 ($\pm$ 80.1) &       51.2 ($\pm$ 73.6) &                        < 0.01 (*)  \\
\bottomrule
\end{tabular}
}
\caption{Overview of clinical variables and their relation to the PCR test result. Data are mean ($\pm$ standard deviation) or n (\%). The significance level of the difference between both groups (positive and negative PCR test) is measured with a Chi-Square test for binary variables and with a Mann-Whitney U test for continuous variables (marked with (*) if < 0.05).
}
\label{tab:variables_overview}
\end{table*}

%% file: figures/table_main_results.tex
\begin{table}[ht]
\centering
\resizebox{\linewidth}{!}{
\begin{tabular}{lcccc}
\toprule
         & \multicolumn{4}{c}{\textbf{Lung Ultrasound}}     \\                                                                    
& Accuracy & F1 & Sensitivity & Specificity  \\ 
\toprule
\multicolumn{4}{l}{\textbf{Human classification}} \\ 
\midrule \midrule
B-lines in $>3$ videos &  0.65  &  0.64 &         0.58 &         0.73  \\
\midrule
\multicolumn{4}{l}{\textbf{CNN-based classification models}} \\ \midrule
\midrule
\textit{Zero-shot (POCOVID-Net)} & 0.65 & 0.66 & 0.62 & 0.70  \\

POCOVID-Net \citep{born2021accelerating}         & 0.73     & \underline{0.78}          & \textbf{0.90}                    & 0.55                              \\ 
ResNet50       & 0.63      & 0.69                   & 0.77                             & 0.50                                            \\ 
NASNetMobile  & 0.58       & 0.72                 & 0.61                             & \textbf{0.90}                        \\ 
MobileNetV2    & 0.68       & 0.68                  & 0.64                             & 0.73                            \\ 
EfficientNetB7 & 0.72      & 0.74                  & 0.77                             & 0.65                             \\ 
\midrule
\multicolumn{4}{l}{\textbf{Pattern-based models + light classifier}} \\ \midrule \midrule
\multicolumn{4}{l}{
\texttt{CovEcho} localization~\cite{josephCovEchoResourceConstrained2022}} \\ 
       + Logistic Regression    &  0.66 & 0.65 &  0.60  &  0.73 \\
       + Random Forest   &  0.54  & 0.60  &  0.66 &  0.42   \\
       + Support Vector Machine   & 0.66 & 0.65 &  0.60  &  0.73    \\
\midrule \multicolumn{4}{l}{
ICLUS segmentation~\cite{roy2020deep}} \\ 
+ Logistic Regression &  \textbf{0.79} & \textbf{0.81} &  \underline{0.82} &  \underline{0.77}  \\
+ Random Forest & \underline{0.74}  & 0.75 &  0.72  &  \underline{0.77}  \\
+ Support Vector Machine & \textbf{0.79} & \textbf{0.81} &  \underline{0.82}  &  \underline{0.77}  \\

\bottomrule
\end{tabular}
}
\caption{Performance of different classification approaches for detecting COVID-19. 
The highest values per column are highlighted bold, the second highest are underlined. A table including the standard deviations over the five folds is given in Appendix F. 
}
\label{tab:CNN_perform}
\end{table}

%% file: figures/table_variables_results.tex
\begin{table*}[ht]
\centering
\resizebox{\textwidth}{!}{
\begin{tabular}{l|ccc|ccc|ccc}
\toprule
&
\textbf{Accuracy} & \textbf{Sensitivity} & \textbf{Specificity} 
& 
\textbf{Accuracy}
& \textbf{Sensitivity} & \textbf{Specificity}&\textbf{Accuracy} & \textbf{Sensitivity} & \textbf{Specificity}\\
\toprule
\multicolumn{4}{l}{\textbf{Random Forest}}                                                                 & \multicolumn{3}{c}{\textbf{Clinical variables}}                                                    & \multicolumn{3}{c}{\textbf{Clinical variables + Blood count}}
\\
\midrule

RF (no images) & - & - & - &          0.63 &        0.60 &        0.67 & \textbf{0.84} &        0.78 &        0.90
\\ 
\midrule \multicolumn{1}{l}{\textbf{CNN classifiers}} & \multicolumn{3}{c}{\textbf{LUS}} & \multicolumn{3}{c}{\textbf{LUS + Clinical variables}}  & \multicolumn{3}{c}{\textbf{LUS + Clinical + Blood count}}  \\ \midrule

POCOVID-Net \citep{born2021accelerating}         & \textbf{0.73}              & \textbf{0.90}                    & 0.55                             & 0.78                        & 0.77                             & 0.80                             & 0.78                        & 0.87                             & 0.69                             \\ 
ResNet50       & 0.63                        & 0.77                             & 0.50                             & \textbf{0.83}               & \textbf{0.84}                    & 0.83                             & 0.80               & 0.67                             & \textbf{0.93}                    \\ 
NASNetMobile  & 0.58                        & 0.61                             & \textbf{0.90}                    & 0.75                        & 0.61                             & \textbf{0.89}                    & 0.79                        & 0.67                             & 0.90                             \\ 
MobileNetV2    & 0.68                        & 0.64                             & 0.73                             & 0.78                        & 0.71                             & 0.87                             & 0.78                        & \textbf{0.88}                    & 0.69                             \\ 
EfficientNetB7 & 0.72                        & 0.77                             & 0.65                             & 0.77                        & 0.81                             & 0.73                             & 0.76                        & 0.64                             & 0.9                              \\ 
\bottomrule
\end{tabular}
}
\caption{Comparison between image-based, variable-based and hybrid approaches. Best performances are marked bold.
}
\label{tab:CNN_perform_variables}
\end{table*}

%% file: appendix.tex
\newpage
\section*{\LARGE{Appendix}}
\renewcommand{\thetable}{A\arabic{table}}  
\renewcommand{\thefigure}{A\arabic{figure}}
\renewcommand{\thesection}{\Alph{section}}
\setcounter{page}{1}
\setcounter{figure}{0}    
\setcounter{table}{0}   
\setcounter{section}{0}

\section{Analysis of labeled severity scores}\label{app:oxygen}

\begin{figure}[!htb]
    \centering
    \includegraphics[width=\columnwidth]{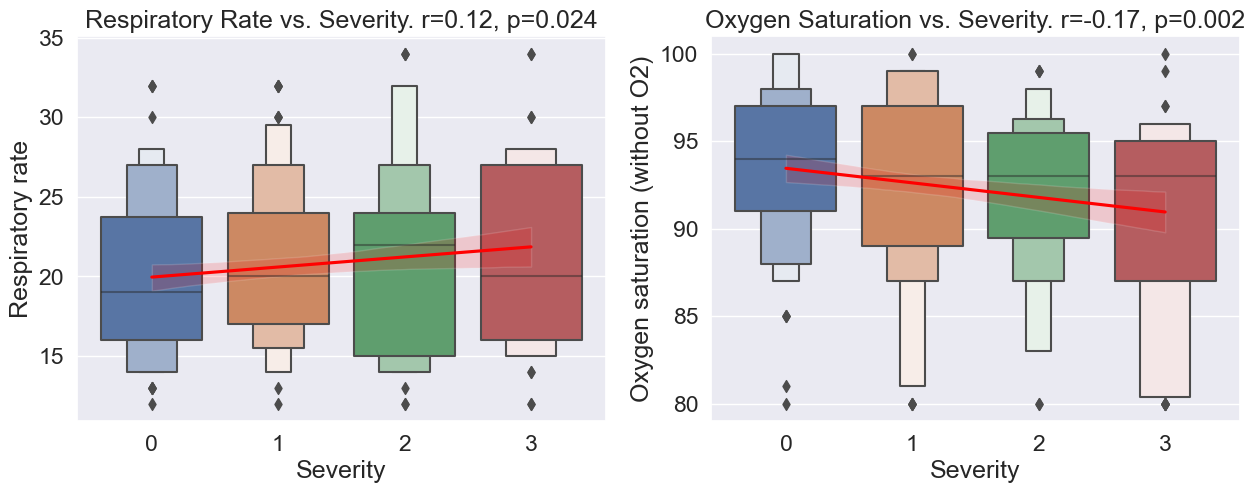}
    \caption[Labeled severity vs Respiratory Rate Oxygen Saturation]{Labeled severity vs Respiratory Rate (left) and Oxygen Saturation (right). The red regression line indicates the correlation between the two variables. Both effects are statistically significant.}
    \label{fig:severity_ox_resp}
\end{figure}

\autoref{fig:severity_ox_resp} shows that the average severity score correlates weakly with respiratory rate and oxygen saturation.

\FloatBarrier
\section{Schemata provided with the CovEcho model}\label{app:covechoSeverity}

\citet{josephCovEchoResourceConstrained2022} devise a scheme to score the frame quality based on the detected landmarks. 30 points are given if at least pleura is detected, 15 and 10 for rib and shadow respectively, and 45 points for either A-lines, B-lines, B patch or consolidations, yielding a maximum of 100.

Furthermore, they provide a scheme to derive a severity score from the detected patterns: 
\setcounter{enumi}{-1} 
\begin{enumerate}[label=\textbf{\arabic*:},font=\normalfont]
    \item[\textbf{0:}]  A-lines are present.
    \item Single or multiple B-lines are detected.
    \item Confluent appearance of B-lines (B patch)
    \item Degraded by the appearance of consolidations due to the effusion in between two pleural surfaces.
    \item Air bronchograms.
\end{enumerate}

\begin{figure}[!htb]
    \centering
    \includegraphics[width=0.5\columnwidth]{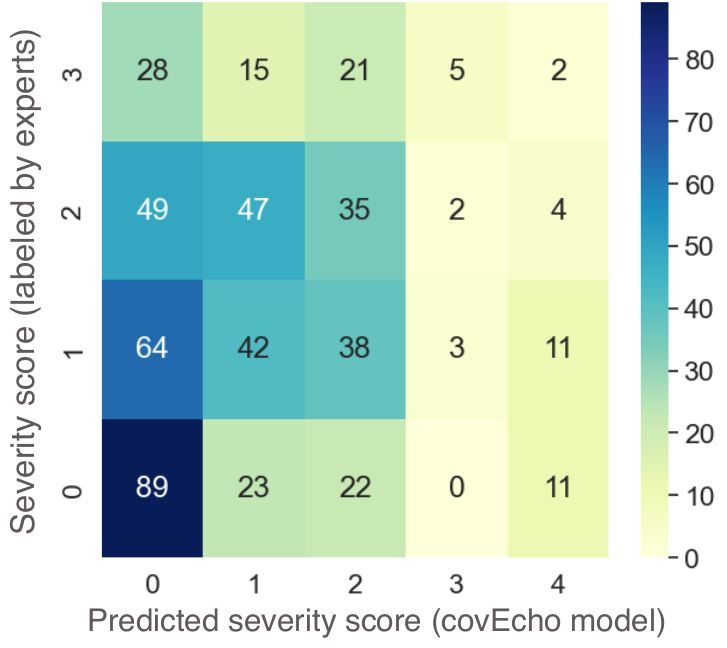}
    \caption{Confusion matrix}
    \label{fig:severity_confusion}
\end{figure}

\autoref{fig:severity_confusion} shows the confusion matrix when comparing our severity scores to the predicted severity scores derived with this scheme. The correlation is low (Spearman $\rho=0.1$) and the CovEcho severity score underestimates the severity score.

\section{Feature importance of clinical variables in RF}\label{app:variables}

\begin{figure}[htb]
    \centering
    \includegraphics[width=\columnwidth]{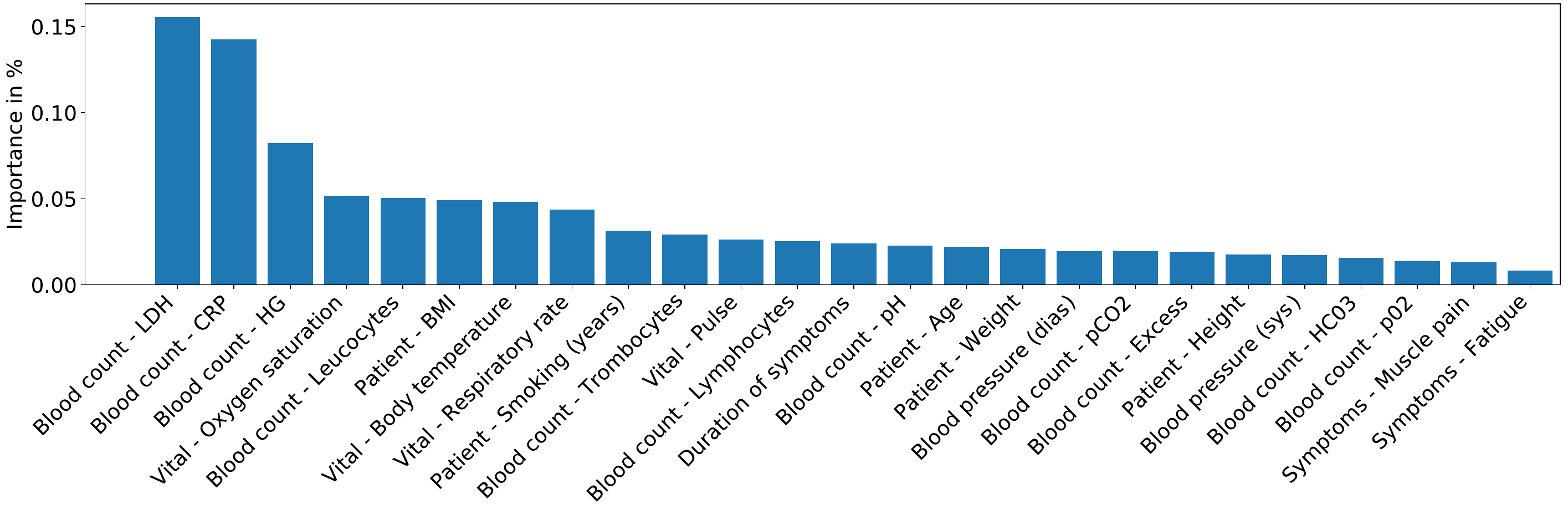}
    \caption{Feature significance of the RF classifier for all features. Shown are only the top 25 most important features of the 66 total.}
    \label{fig:all_feat}
\end{figure}

The RF (Random Forest) model trained to predict the PCR test result solely from clinical variables can be analyzed with feature importance techniques.
Figure \ref{fig:all_feat} visualises the top 25 most important features and reveals that the top four most important features are CBC data. The two most important by a notable margin are lactate dehydrogenase (LDH) and C‐reactive protein (CRP) values. This confirms the results in section~\ref{sec:clinical} that show a substantial decrease in accuracy when witholding blood test data.

\section{Training of convolutional neural networks}\label{app:model_training}

Following~\citep{born2021accelerating}, the models are warm-started from a model pretrained on ImageNet and we only finetune the last six layers. We use Adam optimizer with a learning rate of 0.0001 and cross-entropy loss. 
The hyperparameters and layer sizes were tuned with Weights \& Biases. 
The full network was trained for up to 50 epochs in 5-fold cross validation. 
Furthermore, we found it advantageous to opt for a ``warm start'' of the network by pretraining on the public POCUS dataset by \citet{born2021accelerating} (only including healthy and COVID-19-infected patients); however, this is only possible when training only on images, since the POCUS dataset does not provide sufficient clinical variables. 
Here, we augment data with random rotations (20\% in either direction), random contrast (factor of 0.2) and random zoom to prevent the network from finding a way to classify the images based on the US probe borders.

We further proposed to merge images with clinical variables as input. In this case, the clinical variables are encoded as a normalized vector and passed through one fully-connected layer before being concatenated with the output of the last convolutional layer. The clinical variables provided to the model comprise all variables collected at admission, including the variables listed in \autoref{tab:variables_overview}, together with further variables that are not listed in \autoref{tab:variables_overview} for the sake of simplicity, specifically (for the full table see Supplementary Material):
\begin{itemize}
    \item Inclusion criteria: ``Fever or chills'',``Cough'',``Difficulty breathing'',``Loss of taste'', ``Loss of smell'', ``Sore throat'',``Congestion'',``Runny nose''
    \item Additional collected symptoms: ``Earache'', ``Wheezing'', ``Joint pain'', ``Dyspnea'', ``Decreased consciousness'', ``Confusion'', ``Abdominal pain'', ``Nausea'', ``Vomitting'', ``Diarrhea'', ``Skin rash'', ``Lymphadenopathy'', ``Ageusia/Dysgeusia'', ``Anosmia/Hyposmia'', ``Emergency case''
    \item Pulmonary diseases: ``DPLD'', ``Cystic fibrosis'', ``Pneumothorax'',  ``Tuberculosis'', ``Dyspnea'', ``Other''
\end{itemize}



\section{Severity score prediction}\label{app:confusion}
\begin{figure}[htb]
    \centering
    \begin{subfigure}[b]{0.45\columnwidth}
    \includegraphics[width=\columnwidth]{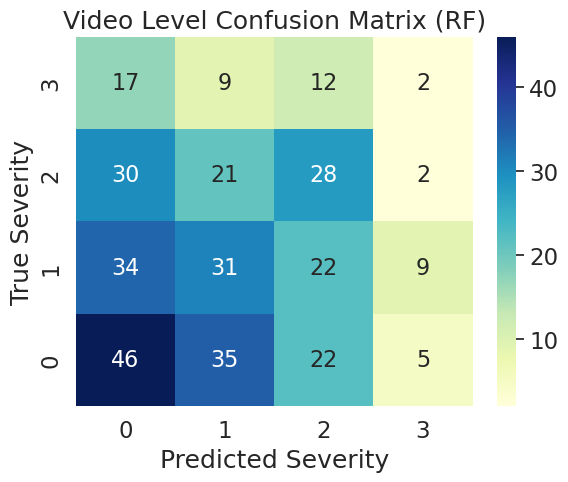}
    \caption{CovEcho-based}
    \end{subfigure}
    \begin{subfigure}[b]{0.45\columnwidth}
    \includegraphics[width=\columnwidth]{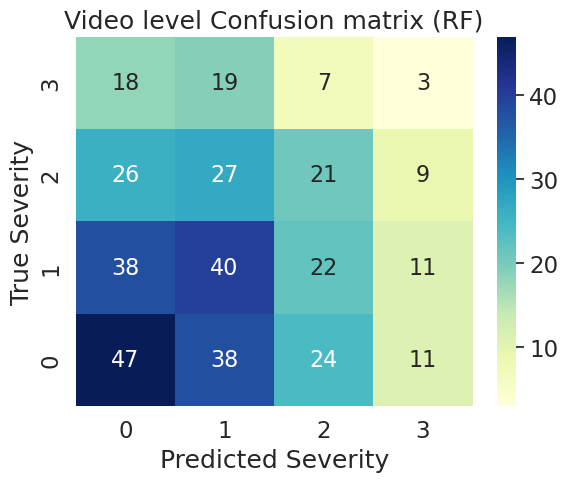}
    \caption{segmentation-based}
    \end{subfigure}
    \caption[Video Level Confusion Matrix]{Confusion matrix of the best performing model for the video level severity score prediction using the CovEcho model class counts and the segmentation class areas respectively.}
    \label{fig:sev_covEcho_confusion_matrix}
\end{figure}

\autoref{fig:sev_covEcho_confusion_matrix} presents the confusion matrices for predicting the disease severity from the CovEcho or ICLUS models. The severity tends to be underestimated.

\section{Results for image-based models for diagnosis}\label{app:cnn_std}
\input{figures/table_std}

\autoref{tab:CNN_perform_std} and \autoref{tab:segmentation_training} list the same results as \autoref{tab:CNN_perform} but including standard deviations.

\input{figures/backup_table}

\section{Relation between segmented pixel count and COVID-19 diagnosis}\label{app:pixel_counts}
\begin{figure}[!bth]
    \centering
    \includegraphics[width=1.0\columnwidth]{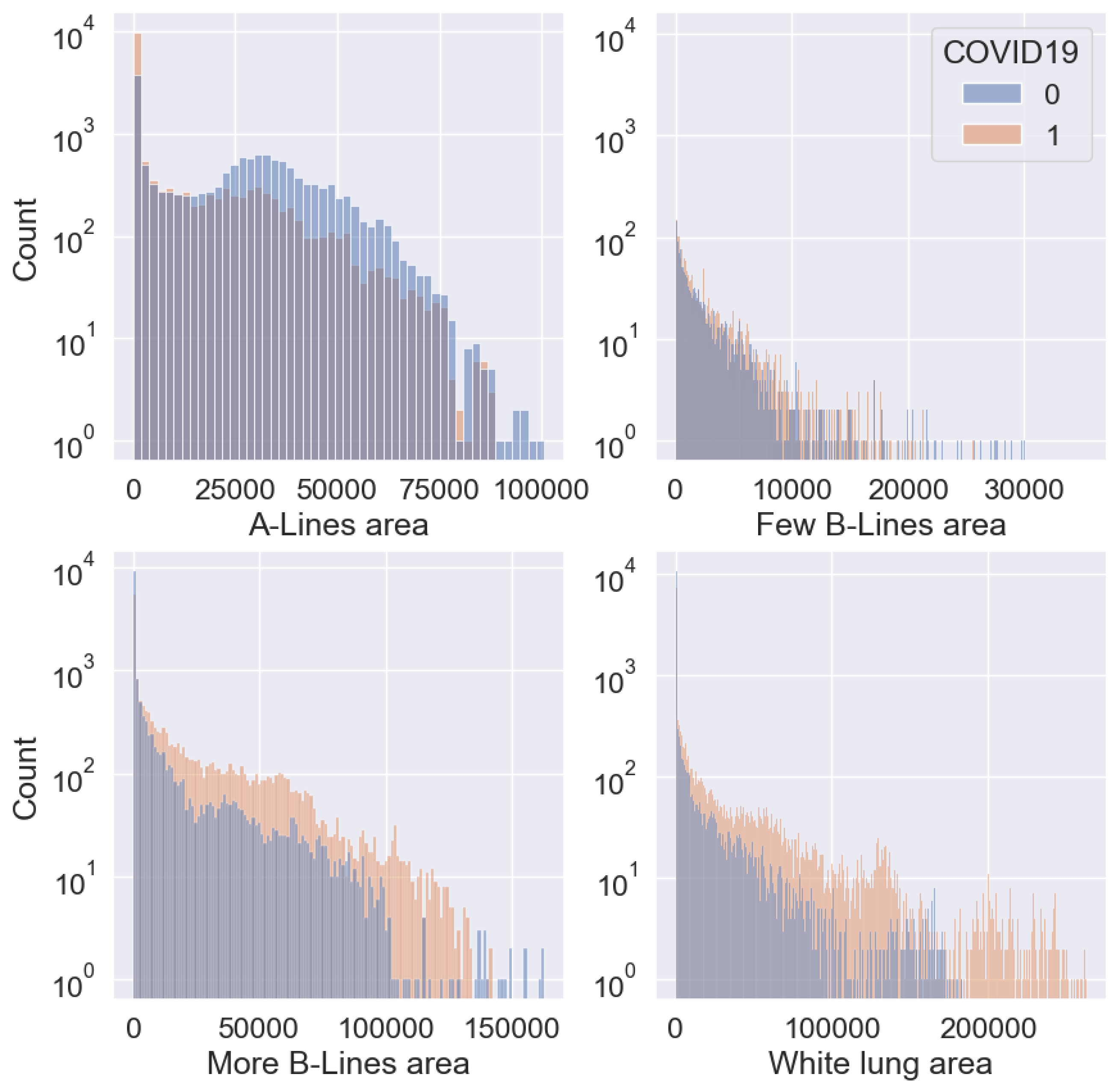}
    \caption[Histogram of pixel counts per frame]{One histogram for each class assigned by the ICLUS model summarizing the pixel counts in all frames. The plots use the logarithmic scale on the y-axis for better visibility. 
    All the distributions are significantly different when split by the COVID-19 variable ($p<0.001$).
    }
    \label{fig:roy_pixel_count_hist}
\end{figure}

\autoref{fig:roy_pixel_count_hist} shows the distribution of pixel count by class. A-lines are detected less frequently for COVID-19 positive cases, whereas B-lines and white lungs are more frequent.

\section{Human diagnosis of COVID-19 from LUS}\label{app:manual_severity}

In \autoref{tab:manual_severity} the results of human diagnosis are given for all tested aggregation schemes. In particular, we tested to average the severity score and to apply different thresholds, or to count the number of videos with B-lines occurring. The highest correspondence to the PCR test result is achieved when assuming that patients with B-lines in more than 3 videos are COVID-positive.

\begin{table}[h!tb]
\centering
\begin{tabular}{llrrr}
\toprule
            Level &  Rule for diagnosis           &  Accuracy &  Sensitivity &  Specificity \\
\midrule
Patient & Average severity $\geq$ 0.5 &      0.49 &         0.85 &         0.10 \\
            & Average severity $\geq$ 1 &      0.54 &         0.58 &         0.50 \\
            & Average severity $\geq$ 2 &      0.49 &         0.18 &         0.83 \\
            & B-lines in > 1 videos &      0.63 &         0.79 &         0.47 \\
            & B-lines in > 2 videos &      0.63 &         0.73 &         0.53 \\
            & \textbf{B-lines in > 3 videos} &      \textbf{0.65} &         0.58 &         0.73 \\
\midrule
Video & B-lines visible &      0.59 &         0.71 &         0.44 \\
            & Severity > 0 &      0.51 &         0.66 &         0.33 \\
            & Severity > 1 &      0.48 &         0.36 &         0.64 \\
\bottomrule
\end{tabular}
\caption{Predicting the PCR test result via manual severity assessment from LUS videos. The best match is achieved with B-line counting}
\label{tab:manual_severity}
\end{table}

%% file: figures/table_std.tex
\begin{table*}[b!ht]
\centering
\resizebox{\textwidth}{!}{
\begin{tabular}{l|l|l|l|l|l|l|l|l|l}
Model          & \multicolumn{3}{c|}{Images}                                                                              & \multicolumn{3}{c|}{Images and Features}                                                        & \multicolumn{3}{c}{\begin{tabular}[c]{@{}c@{}}Images and Features \\ without Blood Test\end{tabular}}  \\ 
\hline
               & Acc.                               & Sens.                    & Spec.                    & Acc.                       & Sens.                   & Spec.                    & Acc.                      & Sens.                    & Spec.                           \\ 
\hline
POCOVID-Net         & \textbf{73.4\%}$_{\pm5}$                      & \textbf{0.90}$_{\pm0.13}$             & 0.55$_{\pm0.24}$                      & 78.4\%$_{\pm9}$                       & 0.87$_{\pm0.12}$                     & 0.69$_{\pm0.19}$                      & 78.4\%$_{\pm8}$                      & 0.77$_{\pm0.14}$                      & 0.80$_{\pm0.12}$                              \\ 
\hline
ResNet50       & 63.2\%$_{\pm14}$                              & 0.77$_{\pm0.33}$                      & 0.50$_{\pm0.35}$                      & \textbf{80.3\%}$_{\pm13}$             & 0.67$_{\pm0.30}$                      & \textbf{0.93}$_{\pm0.08}$             & \textbf{83.4\%}$_{\pm5}$             & \textbf{0.84}$_{\pm0.01}$             & 0.83$_{\pm0.11}$                             \\ 
\hline
NASNetMobile   & 58.2\%$_{\pm7}$                               & 0.61$_{\pm0.25}$                      & \textbf{0.90}$_{\pm0.13}$             & 78.5\%$_{\pm17}$                      & 0.67$_{\pm0.30}$                      & 0.90$_{\pm0.13}$                       & 75.0\%$_{\pm17}$                     & 0.60$_{\pm0.38}$                       & \textbf{0.89}$_{\pm0.09}$                    \\ 
\hline
MobileNetV2    & 68.4\%$_{\pm10}$                              & 0.64$_{\pm0.20}$                       & 0.73$_{\pm0.13}$                      & 78.4\%$_{\pm9}$                       & \textbf{0.88}$_{\pm0.15}$            & 0.69$_{\pm0.22}$                      & 78.3\%$_{\pm13}$                     & 0.70$_{\pm0.20}$                        & 0.87$_{\pm0.12}$                             \\ 
\hline
EfficientNetB7 & 71.7\%$_{\pm9}$                               & 0.77$_{\pm0.17}$                      & 0.65$_{\pm0.25}$                      & 76.3\%$_{\pm10}$                      & 0.64$_{\pm0.20}$                      & 0.89$_{\pm0.09}$                      & 76.8\%$_{\pm9}$                      & 0.81$_{\pm0.12}$                      & 0.73$_{\pm0.27}$                             \\ 
        
\bottomrule
\end{tabular}
}
\caption{Performance of different network architectures on the three configurations of the dataset together with the standard deviation over the five folds. The results of a late fusion between the best three networks can be seen in the last row. Highlighted in bold are the highest values per column.}
\label{tab:CNN_perform_std}
\end{table*}


%% file: figures/backup_table.tex
\begin{table}[!htb]
    \begin{subtable}[t]{0.45\textwidth}
        \centering
        \begin{tabular}{l|ccc}
        \toprule
        Model & Accuracy & Sensitivity & Specificity \\
        \midrule
        LR    & 0.66$_{\pm0.19}$ & 0.60$_{\pm0.16}$ & 0.73$_{\pm0.22}$ \\
        RF    & 0.54$_{\pm0.04}$ & 0.66$_{\pm0.13}$ & 0.42$_{\pm0.17}$ \\
        SVM   & 0.66$_{\pm0.19}$ & 0.60$_{\pm0.16}$ & 0.73$_{\pm0.22}$ \\
        \bottomrule
        \end{tabular}
        \caption{covEcho detection model}
    \end{subtable}
    \hfill
    \vspace{2mm}
    \begin{subtable}[t]{0.45\textwidth}
        \centering
        \begin{tabular}{l|ccc}
        \toprule
        Model & Accuracy & Sensitivity & Specificity \\
        \midrule
        LR    & 0.79$_{\pm0.04}$ & 0.82$_{\pm0.12}$ & 0.77$_{\pm0.19}$ \\
        RF    & 0.74$_{\pm0.04}$ & 0.72$_{\pm0.08}$ & 0.77$_{\pm0.09}$ \\
        SVM   & 0.79$_{\pm0.04}$ & 0.82$_{\pm0.12}$ & 0.77$_{\pm0.19}$ \\
        \bottomrule
        \end{tabular}
        \caption{ICLUS segmentation model}
    \end{subtable}
    \caption{COVID-19 detection accuracy of models trained on segmentation class area (patient-level results).}
    \label{tab:segmentation_training}
\end{table}